\documentclass[sigconf]{acmart}

\usepackage{amsmath,amssymb,amsfonts}
\usepackage{algorithmic}
\usepackage{graphicx}
\usepackage{textcomp}
\usepackage{xcolor}
\usepackage{multicol}
\usepackage{makecell} 
\usepackage{array} 
\usepackage{booktabs}
\usepackage{url}
\usepackage{footmisc}
\usepackage{xspace}
\usepackage{multirow}
\usepackage{tcolorbox}
\usepackage{tablefootnote}
\usepackage[normalem]{ulem}

\newcommand{\rqboxc}[1]{\begin{tcolorbox}[left=3pt,right=3pt,top=3pt,bottom=3pt,colback=gray!35,colframe=gray!35,before skip=5pt,after skip=5pt]#1\end{tcolorbox}}

\newcommand\greybox[1]{%
  \vskip\baselineskip%
  \vspace{-0.8\baselineskip}
  \par\noindent\colorbox{lightgray}{%
    \fontsize{8pt}{11pt}\selectfont

    \begin{minipage}{\linewidth}#1\end{minipage}%
  }%
  \vspace{-0.8\baselineskip}
  \vskip\baselineskip%
}

\usepackage[skip=0.333\baselineskip]{caption}

\newcommand{\emas}{\textit{EM-Assist}\xspace}

\newcommand{\raw}{\textit{RawGPT}\xspace}

\newcommand{\numragrf}{905\xspace}
\newcommand{\rfm}{\textit{RefactoringMiner}\xspace}
\newcommand{\puc}{\textit{PurityChecker}\xspace}
\newcommand{\tool}{\textit{MANTRA}\xspace}
\newcommand{\toolFull}{\textit{\textbf{MANTRA} (\textbf{M}ulti-\textbf{A}ge\textbf{NT} Code \textbf{R}ef\textbf{A}actoring)}\xspace}
\newcommand{\peter}[1]{\textcolor{blue}{{\it [peter says: #1]}}}
\newcommand{\jinqiu}[1]{\textcolor{red}{{\it [Jinqiu says: #1]}}}
\newcommand{\feng}[1]{\textcolor{brown}{{\it [feng says: #1]}}}
\newcommand{\ys}[1]{\textcolor{orange}{{\it [yisen says: #1]}}}

\newcommand{\phead}[1]{\vspace{1mm} \noindent {\bf #1}}

\begin{document}

\title{MANTRA: Enhancing Automated Method-Level Refactoring with Contextual RAG and Multi-Agent LLM Collaboration}
\author{Yisen Xu}
\affiliation{
  \institution{SPEAR Lab, Concordia University}
  \city{Montreal}
  \country{Canada}
}
\email{yisen.xu@mail.concordia.ca}

\author{Feng Lin}
\affiliation{
  \institution{SPEAR Lab, Concordia University}
  \city{Montreal}
  \country{Canada}
}
\email{feng.lin@mail.concordia.ca}

\author{Jinqiu Yang}
\affiliation{
  \institution{O-RISA Lab, Concordia University}
  \city{Montreal}
  \country{Canada}
}
\email{jinqiu.yang@concordia.ca}

\author{Tse-Hsun (Peter) Chen}
\affiliation{
  \institution{SPEAR Lab, Concordia University}
  \city{Montreal}
  \country{Canada}
}
\email{peterc@encs.concordia.ca}

\author{Nikolaos Tsantalis}
\affiliation{
  \institution{Department of Computer Science and Software Engineering, Concordia University}
  \city{Montreal}
  \country{Canada}
}
\email{nikolaos.tsantalis@concordia.ca}

\begin{abstract}
Maintaining and scaling software systems relies heavily on effective code refactoring, yet this process remains labor-intensive, requiring developers to carefully analyze existing codebases and prevent the introduction of new defects. Although recent advancements have leveraged Large Language Models (LLMs) to automate refactoring tasks, current solutions are constrained in scope and lack mechanisms to guarantee code compilability and successful test execution. In this work, we introduce \tool, a comprehensive LLM agent-based framework that automates method-level refactoring. \tool integrates Context-Aware Retrieval-Augmented Generation, coordinated Multi-Agent Collaboration, and Verbal Reinforcement Learning to emulate human decision-making during refactoring while preserving code correctness and readability. Our empirical study, conducted on 703 instances of “pure refactorings” (i.e., code changes exclusively involving structural improvements), drawn from 10 representative Java projects, covers the six most prevalent refactoring operations. Experimental results demonstrate that \tool substantially surpasses a baseline LLM model (\raw), achieving an 82.8\% success rate (582/703) in producing code that compiles and passes all tests, compared to just 8.7\% (61/703) with \raw. Moreover, in comparison to IntelliJ’s LLM-powered refactoring tool (\emas), \tool exhibits a 50\% improvement in generating \textit{Extract Method} transformations. A usability study involving 37 professional developers further shows that refactorings performed by \tool are perceived to be as readable and reusable as human-written code, and in certain cases, even more favorable. These results highlight the practical advantages of \tool and emphasize the growing potential of LLM-based systems in advancing the automation of software refactoring tasks.
\end{abstract}

\maketitle

\section{Introduction}


Refactoring is the process of improving the overall design and structure of the code without changing its overall behavior~\cite{DBLP:books/daglib/0019908}. The goal of refactoring is to improve maintainability and facilitate future functionality extension~\cite{murphy2011we}, making it essential for adapting to ever-evolving software requirements. 
However, despite its benefits, many developers hesitate to refactor due to the time and effort involved~\cite{DBLP:journals/infsof/RebaiKASK20}. Developers often need to first analyze the possibility of refactoring, then modify the code to refactor, and finally, ensure that refactoring does not introduce new issues.

To assist developers with refactoring, researchers and Integrated Development Environment (IDE) developers (e.g., Eclipse and Intellij IDEA) have proposed automated refactoring techniques. For example, ~\citet{DBLP:journals/tse/TsantalisC09, DBLP:journals/jss/TsantalisC11} proposed JDeodorant to detect code smells, such as \textit{Feature Envy} and \textit{Long Method}, and apply refactoring operations. 
WitchDoctor ~\cite{foster2012witchdoctor} makes refactoring recommendations by monitoring whether code changes trigger predefined specifications. 
One common characteristic of these tools is that they are based on pre-defined rules or metrics. 
Although useful, they lack a deep understanding of the project's domain-specific structure and cannot produce refactorings similar to those written by developers, resulting in 
low acceptance in actual development~\cite{DBLP:conf/sigsoft/SilvaTV16, 10.1145/3487062}.

Recent research on Large Language Models (LLMs) has shown their great potential and capability in handling complex programming tasks~\cite{DBLP:conf/kbse/WuM0G0ZZ024, DBLP:journals/pacmse/WadhwaPSSNKPR24, ye2025llm4effileveraginglargelanguage, lin2024soen101codegenerationemulating}, making them a possible solution for overcoming prior challenges, (i.e., generating high-quality refactored code similar to human-written ones). Several studies~\cite{murphy2011we,depalma2024exploring, poldrack2023ai,alomar2024refactor,shirafuji2023refactoring} have already explored the use of LLMs for refactoring, demonstrating their strong ability to analyze refactoring opportunities and applying code changes. 
However, existing techniques primarily rely on simple prompt-based refactoring generation, focus on a limited set of refactoring types, and lack proper verification through compilation checks and test execution. Moreover, these approaches have not fully utilized the self-reflection \cite{DBLP:conf/nips/ShinnCGNY23} and self-improvement capabilities of large language models, resulting in limited effectiveness and performance that has yet to match human-level proficiency in code refactoring.

In this paper, we propose \toolFull, an end-to-end LLM agent-based solution for automated method-level refactoring. Given a method to refactor and a specified refactoring type, \tool generates fully compilable, readable, and test-passing refactored code. \tool includes three key components: (1) \textbf{\textit{Context-Aware Retrieval-Augmented Refactored Code Generation}}, which constructs a searchable database to provide few-shot examples for improving refactoring quality; (2) \textbf{\textit{Multi-Agent Refactored Code Generation}}, which employs a Developer Agent and Reviewer Agent to simulate real-world refactoring processes and produce high-quality refactoring; and (3) \textbf{\textit{Self-Repair Using Verbal Reinforcement Learning}}, which iteratively identifies and corrects issues that cause compilation or test failures using a verbal reinforcement learning framework~\cite{DBLP:conf/nips/ShinnCGNY23}.

We evaluated \tool using 10 Java projects used in prior refactoring studies~\cite{DBLP:conf/icse/GrundCBHH21, 10.1145/3540250.3549079, Hasan:TSE:2024:CodeTracker2.0}. These projects cover diverse domains, have rich commit histories, and contain many tests. 
Since most refactoring changes are accompanied by unrelated code changes (e.g., bug fixes or feature additions) \cite{DBLP:conf/sigsoft/SilvaTV16}, we collected ``pure refactoring changes'' (i.e., no code changes other than refactoring) to eliminate noise when evaluating \tool's refactoring ability. 
We applied \puc~\cite{nouri2023puritychecker} to filter commits, ultimately obtaining 703 pure refactorings for our experiments, which cover six of the most common refactoring activities~\cite{golubev2021thousandstorieslargescalesurvey,DBLP:conf/sigsoft/SilvaTV16, Peruma_2021}: \textit{Extract Method}, \textit{Move Method}, \textit{Inline Method}, along with related compound refactoring activities: \textit{Extract and Move Method}, \textit{Move and Inline Method}, and \textit{Move and Rename Method}. Using these refactorings, we compare \tool with our LLM baseline (\raw), IntelliJ’s LLM-based refactoring tool (\emas~\cite{EM-Assist0.7.5}), and human-written refactored code. Furthermore, we conducted a user study to receive developer feedback on \tool-generated code and an ablation study to evaluate the contribution of each component within \tool.

Our results demonstrate that \tool outperforms \raw in both functional correctness and human-likeness (how similar it is compared to developer refactoring). \tool achieves a significantly higher success rate of 82.8\% (582/703) in generating compilable and test-passing refactored code, compared to only 8.7\% for \raw. 
Against \emas, \tool shows a 50\% improvement in generating \textit{Extract Method} refactorings. Compared to human-written code, our user study (with 37 developers) found that \tool and human refactorings share similar readability and reusability scores. 
However, \tool performs better in \textit{Extract \& Move} and \textit{Move \& Rename} refactorings due to its clear comments and better code naming. In contrast, humans do better in \textit{Inline Method} refactoring by making additional improvements. 
Finally, our ablation study highlights that removing any component from \tool results in a noticeable performance drop (40.7\% - 61.9\%), with the Reviewer Agent contributing the most to overall effectiveness.

We summarize the main contributions as follows:
\begin{itemize}
\item We proposed an end-to-end agent-based refactoring solution \tool, which considers compilation success and functional correctness in the refactoring process. \tool leverages Context-Aware Retrieval-Augmented Generation to learn developer refactoring patterns, integrates multiple LLM agents to simulate the developer's refactoring process, and adopts a verbal reinforcement learning framework to improve the correctness of the refactored code.  

\item We conducted an extensive evaluation, and \tool successfully generated 582/703 compilable and test-passing refactorings, significantly outperforming \raw, which only produced 61 successful refactorings. Compared with \emas, a state-of-the-art LLM-based technique primarily focused on \textit{Extract Method} refactoring, \tool achieved a 50\% improvement.

\item 
We conducted a user study to compare
\tool’s generated and human-written refactored code. The analysis of 37 responses shows that the refactored code generated by the \tool is similar to the developer-written code in terms of readability and reusability. Moreover, \tool’s generated code has better advantages in method naming and code commenting.

\item We made the data and code publicly available~\cite{muarf_data_code}.
\end{itemize}

\noindent{\textbf{Paper Organization}.} Section \ref{sec:related} discusses related work. Section \ref{sec:methodology} details the design and implementation of \tool. Section \ref{sec:evaluation} presents the evaluation results of \tool. Section \ref{sec:threats} discusses the limitations and potential threats to validity. Finally, Section \ref{sec:conclusion} summarizes our findings and outlines directions for future work.

\section{Related Work} \label{sec:related}

\noindent{\textbf{Traditional Refactoring Approaches}.}
Refactoring plays a critical role in software development and greatly influences software quality. Traditional research in refactoring generally focuses on two main aspects: \textit{identifying refactoring opportunities} and \textit{recommending refactoring solutions}. In terms of opportunity identification, existing approaches have explored various methods, such as calculating distances between entities and classes for \textit{Move Method} refactoring~\cite{DBLP:journals/tse/TsantalisC09}, assessing structural and semantic cohesion for \textit{Extract Class} opportunities ~\cite{DBLP:journals/jss/BavotaLO11}, and utilizing logic meta-programming techniques to uncover refactoring possibilities~\cite{DBLP:conf/csmr/TourweM03}. On the other hand, solution recommendation often involves automated techniques.  ~\cite{DBLP:journals/jss/OKeeffeC08} 
introduced a tool CODe-Imp, which uses search-based techniques to perform refactoring. WitchDoctor ~\cite{foster2012witchdoctor} and BeneFactor ~\cite{ge2012reconciling}  monitor code changes and automatically suggest refactoring operations. Nevertheless, these approaches are often limited by their rule-based nature, which may restrict them to certain types of refactoring or cause them to encounter unhandled issues during the refactoring process.

\noindent{\textbf{LLM-Based Techniques for Generating Refactored Code}.} 
Recent research on LLMs has demonstrated their remarkable ability to handle complex tasks, making them promising solutions to overcome the limitations of traditional refactoring approaches. Existing studies utilize LLMs for various refactoring tasks: directly prompting GPT-4 for refactoring tasks ~\cite{depalma2024exploring, poldrack2023ai}, providing accurate identification of refactoring opportunities through carefully designed prompts~\cite{liu2024empirical}, and recommending specific refactoring types such as \textit{Extract Method}~\cite{silva2014recommending}. Other studies further improve LLM-based refactoring by emphasizing prompt clarity~\cite{alomar2024refactor}, using carefully selected few-shot examples~\cite{shirafuji2023refactoring}, and applying structured prompting techniques~\cite{white2024chatgpt}. Additionally, hybrid approaches combining rule-based systems and LLMs have achieved superior outcomes compared to single-method techniques ~\cite{zhang2024refactoring}. Automated frameworks and tools, such as the Context-Enhanced Framework for Automatic Test Refactoring~\cite{gao2024context} and tools like \emas~\cite{pomiannext}, further illustrate practical implementations of these techniques, highlighting the value of integrating LLMs into comprehensive, feedback-driven refactoring workflows. \emas even outperforms all prior tools on \textit{Move Method} refactoring~\cite{pomiannext}. 

While these techniques leverage LLMs for automated refactoring, they typically focus on only one or two types of refactoring, neglecting compound or repository-level transformations (e.g., \textit{Extract and Move Method}) and failing to ensure that the refactored code compiles and passes all tests. In contrast, \tool uses LLM agents to emulate developers' refactoring process and integrate traditional tools to provide feedback. \tool generates refactored code for a broader range of refactoring activities and ensures the generated code can compile and pass all tests. 

\noindent{\textbf{LLM-Based Approaches for Code Quality Improvement. }} 
Previous studies have also explored using LLM to improve other aspects of software quality, such as security, performance, and design. For instance, ~\citet{lin2024soen101codegenerationemulating} proposed leveraging Software Process Models to enhance the design quality in code generation tasks. ~\citet{ye2025llm4effileveraginglargelanguage} introduced LLM4EFFI, conducting comprehensive research on improving code efficiency.  ~\citet{DBLP:journals/pacmse/WadhwaPSSNKPR24} proposed CORE, an approach utilizing instruction-following LLMs to assist developers in addressing code quality issues through targeted revisions.  ~\citet{DBLP:conf/kbse/WuM0G0ZZ024} presented iSMELL, which integrates LLMs for the detection and subsequent refactoring of code smells, thus systematically enhancing software quality. 
Inspired by these studies, we designed \tool to also consider code readability by integrating with code style checkers (i.e., CheckStyle~\cite{checkstyle}) in the generation process. 
\section{Methodology} \label{sec:methodology}  

In this section, we introduce \toolFull, an LLM-based, agent-driven solution for automated code refactoring. 
\tool focuses on method-level refactorings because of their wide adoption in practice
~\cite{Negara:2013, kim2014empirical}. In particular, we implement a total of six refactoring activities, composing three of the most popular refactoring activities~\cite{golubev2021thousandstorieslargescalesurvey,DBLP:conf/sigsoft/SilvaTV16, Peruma_2021}: \textit{Extract Method}, \textit{Move Method}, and \textit{Inline Method}; and three of their compound refactoring activities: \textit{Extract And Move Method}, \textit{Move And Inline Method}, and \textit{Move And Rename Method}. These refactoring activities consider both straightforward and intricate refactoring scenarios.

\tool takes as input the code of a method to be refactored and the specified refactoring type. It then automatically finds refactoring opportunities in the method and generates fully compilable and highly readable refactored code that can pass all the tests. 
\tool consists of three key components: 1) \textit{RAG}: Context-Aware Retrieval-Augmented Refactored Code Generation, 2) \textit{Refactored Code Generation}: Multi-Agent Refactored Code Generation, and 3) \textit{Repair}: Self-Repair Using Verbal Reinforcement Learning. The \textit{RAG} component constructs a searchable database that contains prior refactorings as few-shot examples to guide \tool. The \textit{Refactored Code Generation} component uses a multi-agent framework that harnesses LLMs' planning and reasoning abilities to generate refactored code. Finally, the \textit{Self-Repair} component implements a verbal reinforcement learning framework~\cite{DBLP:conf/nips/ShinnCGNY23} to automatically identifies and corrects issues in the generated refactored code. 

\subsection{{Context-Aware Retrieval Augmented Refactored Code Generation}}

\label{subsec:context_rag}

\begin{figure*}
  \centering
  \scalebox{0.95}{
  \includegraphics[width=1\textwidth]{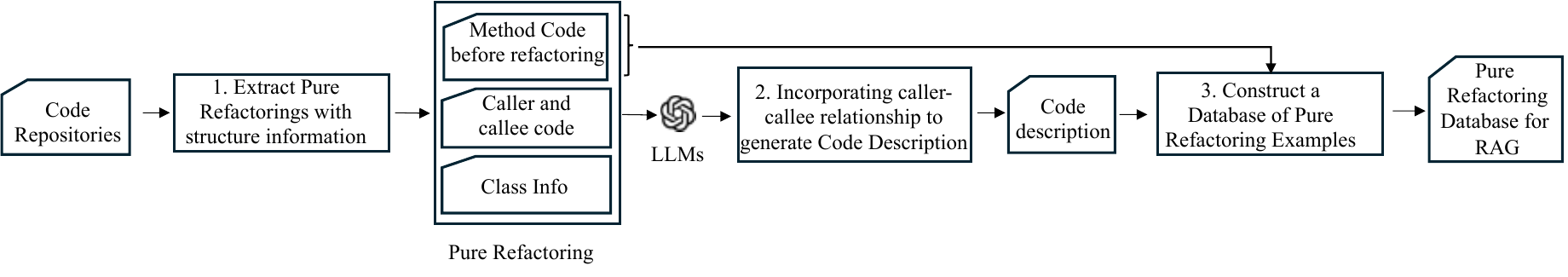}}
  \caption{An overview of how \tool constructs a database containing only pure-refactoring for RAG.}
  \label{fig:rag}
\end{figure*}

Figure~\ref{fig:rag} shows an overview of \tool's RAG construction process. RAG provides LLM with relevant examples for few-shot learning, thus improving its ability to generate accurate and contextually relevant output \cite{DBLP:journals/corr/abs-2410-09662, shirafuji2023refactoring}. RAG combines information retrieval and text generation by integrating external knowledge through a two-stage process -- retrieval and generation \cite{DBLP:journals/corr/abs-2312-10997}. 
Below, we discuss the details of \tool's RAG design. 

\phead{Constructing a Database of Pure Refactoring Code Examples.} 
We aim to build a database containing only pure refactoring code changes, which involve improving code structure without altering functionality. 
However, in reality, refactoring is often accompanied by unrelated code changes such as bug fixes or feature additions \cite{DBLP:conf/sigsoft/SilvaTV16}. These changes contain noise that makes such refactoring code changes unusable as few-shot examples to guide LLM for generating general refactored code.

We build the pure refactoring database using the \textit{Refactoring Oracle Dataset} \cite{tsantalis2020refactoringminer}, which contains 12,526 refactorings (mostly impure refactorings) collected from 547 commits across 188 open-source Java projects (2011 to 2021). 
We selected this dataset because it includes various projects and refactoring types and was used as a benchmark to evaluate the accuracy of refactoring detection tools (e.g., \textit{RefDiff}~\cite{RefDiff2.0:2021} and \textit{RefactoringMiner}~\cite{tsantalis2018accurate, tsantalis2020refactoringminer}), making it highly suitable for our purpose. 
We incorporated \puc~\cite{nouri2023puritychecker}, an extension of \rfm that specializes in assessing the purity of method-level refactorings, excluding those associated with functional feature modifications. We choose to use \rfm and \puc because they are well-maintained and known for their high detection accuracy. \rfm has an average precision and recall of 99\% and 94\%, respectively~\cite{tsantalis2020refactoringminer} and \puc has an average precision and recall of 95\% and 88\%, respectively~\cite{nouri2023puritychecker} on the \textit{Refactoring Oracle Dataset}. 
At the end of this phase, we extracted \numragrf pure refactorings along with their associated metadata (e.g., \textit{Class Name}, \textit{Method Signature}, \textit{File Path}, and \textit{Call Graph}) from the GitHub repositories. 

\phead{Incorporating Code Description and Caller-Callee Relationships for Context-Aware RAG Retrieval. }
Using only source code to construct a RAG database presents several challenges. First, code with similar structures does not necessarily share the same functionality or logic, both of which influence refactoring strategies. For instance, in \textit{Move Method} refactoring, a test method should only be moved to a test class, not to production code. If the context does not clearly indicate the class type, relying solely on source code structure for retrieval may result in incorrect matches. 
Second, code dependencies play a crucial role in refactoring, as refactorings like \textit{Extract Method} and \textit{Move Method} are often driven by code dependencies \cite{DBLP:conf/cascon/TsantalisGSH13}. Without capturing these dependencies, the retrieved examples may fail to align with the intended refactoring process.

Therefore, in \tool, we incorporate 1) a natural language description of the refactored code and 2) all direct callers and callees of the refactored method as code-specific context to enhance RAG’s retrieval capabilities. To generate the natural language description, we follow a recent study in the NLP community~\cite{contextual-retrieval}. We use LLMs to generate a contextual description for every refactoring and concatenate this description with the corresponding code to construct the contextual database. 
To guide the LLM in generating these descriptions, we use a simple prompt: \textbf{\tt ``\{Code\}\{Caller/Callee\}\{Class Info\}
Please give a short, succinct description to situate this code within the context 
to improve search retrieval of the code.''}, where {\tt\{Code\}} refers to the code before refactoring, {\tt \{Caller/Callee\}} denotes the direct callers/callees, and {\tt\{Class Info\}}  presents the structure information for the Class containing the code to be refactored, such as \textit{Package Name}, \textit{Class Name}, \textit{Class Signature}.
The prompt includes the direct callers and callees of method to be refactored to assist in description generation, as dependent methods may influence refactoring decisions. Specifically, the prompt contains the method signatures and bodies of all direct callers and callees, enabling the retrieval mechanism to account for such dependencies.  

\phead{Retrieving the Most Similar Code As Few-Short Examples.} 
In the retrieval stage of RAG, two main retrieval methods are commonly combined and fused for the best outcome~\cite{DBLP:journals/corr/abs-2410-09662}, namely sparse retrieval and dense trieval. Sparse retrieval uses textual similarity to efficiently retrieve relevant documents, and dense retrieval relies on semantic similarity.
\tool leverages sparse and dense retrieval separately for producing two similarity-ranked lists, then combines the lists to create a unified ranking list. 
For sparse retrieval, \tool leverages BM25 \cite{robertson2009probabilistic}, a robust ranking technique to obtain a ranked list based on textual similarity. For dense retrieval, \tool employs \textit{all-MiniLM-L6-v2}, a pre-trained model from Sentence Transformers \cite{reimers-2019-sentence-bert} known for its speed and quality, for embedding generation. 
\tool then computes the cosine similarity between embeddings of the input refactoring request and stored refactoring examples to obtain an additional ranked list based on semantic similarity.
Finally, \tool uses the Reciprocal Rank Fusion (RRF) algorithm ~\cite{DBLP:conf/sigir/CormackCB09, DBLP:conf/naacl/SanthanamKSPZ22} to combine the sorted lists and re-rank the results. 
\subsection{Multi-Agent Refactored Code Generation} 
\begin{figure*}

  \includegraphics[width=1\textwidth]{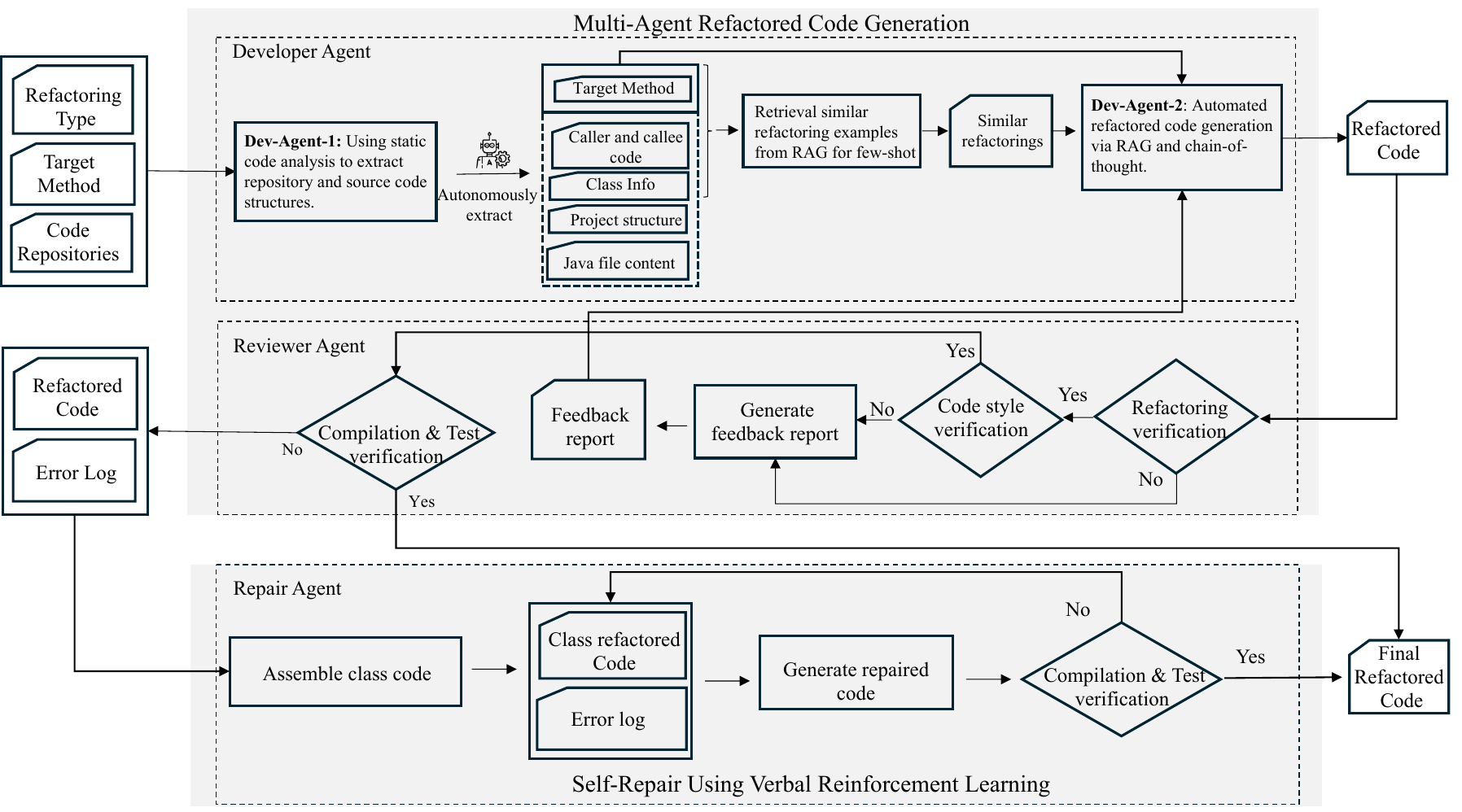} \centering
  \caption{An overview of \tool. }
  \label{fig:methodology}
\end{figure*}

\tool emulates how real-world code refactoring happens through a multi-agent collaboration among the Developer Agent, the Reviewer Agent, and the Repair Agent. As shown in Figure~\ref{fig:methodology}, \tool adapts a mixture-of-agents architecture~\cite{DBLP:journals/corr/abs-2406-04692} to organize the agent communication in two layers. In the first layer, the Developer Agent is responsible for generating and improving the code, while the Reviewer Agent reviews the code and provides feedback or suggestions to the Developer Agent. 
If the refactored code fails to compile or pass tests, the generated code 
enters the second layer of the agent architecture, where the Repair Agent tries to repair any compilation or test failures based on LLM-based reasoning and verbal reinforcement learning~\cite{DBLP:conf/nips/ShinnCGNY23}.


\subsubsection{Developer Agent}

Given a method to refactor and a specified refactoring type, the Developer Agent first autonomously extracts the necessary information (e.g., repository structure, class information) based on the observation (i.e., the refactoring type and the provided inputs) by invoking our custom static analysis. It then retrieves similar refactorings using our contextual RAG approach (Section 3.1) as few-shot examples to enhance code generation. Finally, it generates the refactored code using the extracted information and the retrieved examples.

\phead{Dev-Agent-1: Using static code analysis to extract repository and source code structures.} 
The Developer Agent has access to our custom static analysis tools to analyze the repository and extract code and project structural information. The code structural information includes the class hierarchies, inheritance relationship, method signatures and their implementation in a class, and interprocedural method call graph. The project structural information includes the project directory structure and the specific Java file content. Among these, method signatures and their implementation are mandatory for all refactoring types, whereas other information is only necessary for specific types of refactoring. 
 
 To reduce static analysis overhead and unneeded information to the LLM, the Developer Agent autonomously decides which analyses to perform based on the given refactoring type and the target code. For example, for the \textit{Move Method} refactoring, the Developer Agent first calls \textit{get\_project\_structure} to retrieve the overall project structure. Based on this information, it determines the relevant file directories to inspect. It then calls \textit{get\_file\_content} to retrieve the source code files from the directory and assesses whether the method should be moved to the target class/file. The Developer Agent then uses the analysis results in the next phase to guide the refactoring process and generate the refactored code.

\phead{Dev-Agent-2: Automated refactored code generation via RAG and chain-of-thought.} 
Given the static analysis result from the previous step, the Developer Agent 1) retrieves similar refactorings as few-shot examples using RAG and 2) generates the refactored code using chain-of-thought reasoning~{\cite{wei2023chainofthoughtpromptingelicitsreasoning}}. The Developer Agent provides the code to be refactored, the generated code description, and direct callers/callees as input to the RAG database to retrieve similar refactorings for few-shot learning. 
Then, the Developer Agent follows a structured chain-of-thought reasoning approach, analyzing the provided information sequentially. 
Below, we provide a \textbf{\textit{simplified}} prompt example to show the code generation process. 
\greybox{{\textit{
\textbf{\#\#\#Task: }Code Refactoring Based on a Specified Refactoring Type
\\
\\
\textbf{\#\#\#Instructions: }Please follow the Step-by-Step Analysis:
\\
\textbf{Step 1: Code Analysis.} Analyze the specific code segment that needs to be refactored. And output a concise summary of the code to be refactored.
\\
\textbf{Step 2: Refactoring Method Reference.} Search and retrieve up to three similar refactoring examples from the RAG system.
\\
\textbf{Step 3: Structure Information Extraction.} Based on the refactoring type and code summary, use the provided tools to collect any structural information you need. This may include code structure information as well as project structure information.
\\
\textbf{Step 4: Refactoring Execution.} Using the extracted structural information and retrieved examples to generate the refactored code.
\\
\\
\textbf{\#\#\# Input: }
`\{Code to be Refactored, Refactoring Type\}'
\\
\textbf{\#\#\# Response:} 
`\{Refactored Code\}'
}}}

As shown in the example, the Developer Agent analyzes the method source code and the entire class in which it resides.
Then, the agent autonomously decides whether to further analyze broader contextual information, including direct callers/callees, the inheritance graph, and the repository structure. 
After collecting source code and other structural information, the agent then generates a list of potential refactoring opportunities for the given refactoring type, such as which parts of a method can be extracted (e.g., for \textit{Extract Method}) or possible classes to move a method to (e.g., for \textit{Move Method}). 
Finally, the agent generates the refactored code based on the retrieved few-shot examples and the list of potential refactoring opportunities.
Finally, the agent selects the most probable refactoring opportunity from the list and generates the refactored code based on the retrieved few-shot examples.

\subsubsection{Reviewer Agent}
The Reviewer Agent is responsible for evaluating the refactored code generated by the Developer Agent, ensuring its correctness in two key aspects: refactoring verification and code style consistency to ensure readability. 
The Reviewer Agent first verifies whether the code has undergone the specified type of refactoring by leveraging \rfm to detect refactoring activities in the code. 
If the code fails this check, the Reviewer Agent immediately generates a feedback report containing the refactoring verification results and returns it to the Developer Agent for correction. If the refactoring verification is successful, the Reviewer Agent proceeds with code style consistency analysis, which is conducted using the static analysis tool CheckStyle \cite{checkstyle}. If any code style issues are detected, such as non-standard variable naming or formatting inconsistencies, the Reviewer Agent generates a feedback report highlighting these problems and sends it back to the Developer Agent for correction. 
Once the refactored code passes the refactoring verification and code style consistency check, the Reviewer Agent will compile and test it. If there is no failure and all tests pass, the generation process is done, and \tool returns the final refactored code. If there is any failure, the generated code and error log will be forwarded to the next phase.

\subsubsection{Communication between the Developer and Reviewer Agents}

In \tool, the Developer Agent and the Reviewer Agent work together in an iterative process, simulating human team collaboration. After generating refactored code, the Developer Agent submits it for review. The Reviewer Agent then verifies both the refactoring and code style, providing structured feedback. If any issues are found, the Developer Agent refines and resubmits the code, repeating the cycle until all required standards are met.

After verifying the refactoring activity and code style, the Reviewer Agent triggers the compilation and testing phase. 
The agent integrates the generated refactored code into the project and creates compilation commands based on the project's build system (Maven or Gradle). It then executes the commands to compile the project and run the tests. If there are any compilation issues or test failures, the issues are escalated to the Repair Agent for fixing.




\subsection{Self-Repair Using Verbal Reinforcement Learning}

The Repair Agent iteratively fixes refactored code that fails to compile and pass tests by leveraging verbal reinforcement learning. To enhance its self-repairing capability, we integrate and adapt the Reflexion framework~\cite{DBLP:conf/nips/ShinnCGNY23}, which systematically guides the repair process through four distinct phases: initial analysis, self-reflection, planning, and acting. Reflexion is designed to enable systems to self-improve by incorporating iterative feedback.

 The Repair Agent starts an \textit{(1) initial analysis}, where the Repair Agent examines the refactored code, the entire class where the code resides, and associated error logs. 
 Then, it generates an initial patch based on the problematic code segments and corresponding error descriptions. 
 Subsequently, the agent applies the patch, re-compiles the code, and re-runs the tests. If there are any issues, the agent enters the \textit{(2) self-reflection} phase, critically self-reviewing the compilation and testing result. 
 The agent generates error reasoning by comparing the code and the associated error messages before and after applying the patch. 
 For instance, if the previous patch fails to resolve a null pointer exception, the agent reflects on the absence or inadequacy of necessary null checks by explicitly referencing the corresponding lines in the stack trace. 

 Based on the self-reflection result, the Repair Agent enters the \textit{(3) planning} phase, where it generates a refined repair strategy, specifying concrete code modifications required to resolve identified issues (e.g., adding necessary null checks, correcting variable declarations, or revising incorrect method calls). Finally, in the \textit{(4) acting} phase, the Repair Agent applies the planned patch, followed by compilation and test execution. 
 This is an iterative process that continues the code successfully compiles and passes all tests or reaches a predefined maximum number of repair attempts (i.e., 20). To ensure the code semantic remains unchanged, during the repair process, we specify in the prompt that the agent should not modify the code's functionality and only focus on repair.

\begin{table}
    \caption{The studied Java projects for evaluating \tool. }
    \label{tab:proj_info}
    \centering
     \setlength{\tabcolsep}{4mm}
    \scalebox{0.8}{
        
        \begin{tabular}{lrrr}		
            \hline
            \textbf{Project} & \textbf{\# Star} & \textbf{\# Commits} & \textbf{\# Pure Refactoring} \\
            \hline
checkstyle	&	8,462	&	14,606	&	91	\\
pmd	&	4,988	&	29,117	&	125	\\
commons-lang	&	2,776	&	8,404	&	59	\\
hibernate-search	&	512	&	15,716	&	89	\\
junit4	&	8,529	&	2,513	&	18	\\
commons-io	&	1,020	&	5,455	&	93	\\
javaparser	&	5,682	&	9,607	&	56	\\
junit5	&	6,523	&	8,990	&	105	\\
hbernate-orm	&	6,091	&	20,638	&	63	\\
mockito	&	15,032	&	6,236	&	4	\\
    \hline
\textbf{Total}	&	59,615	&	121,282	&	703	\\	
\hline
        \end{tabular}
    }
       
\end{table}

\section{Evaluation} 
In this section, we first present the studied dataset and evaluation metrics. Then, we present the answers to the four research questions. 

\label{sec:evaluation}


\phead{\textbf{Studied Dataset}.} Table \ref{tab:proj_info} shows the Java projects that we use to evaluate \tool. 
We select these 10 Java projects used in prior change history tracking studies~\cite{DBLP:conf/icse/GrundCBHH21, 10.1145/3540250.3549079, Hasan:TSE:2024:CodeTracker2.0} based on three key considerations. First, the projects cover diverse application domains, providing a broad representation of software development practices. Second, each project has a substantial commit history, with over 2,000 commits, indicating a richer development history and a higher likelihood of identifying commits that involve refactoring activities. Third, we chose projects where we could manually resolve the compilation issues and successfully execute the tests to verify the quality of the generated refactoring. 

We evaluate and compare the refactored code generated by \tool with that produced by human developers. To reduce noise from unrelated changes, such as bug fixes, we analyze only ``pure refactoring changes'' (i.e., no code changes other than refactoring) from these projects. Similar to Section \ref{subsec:context_rag}, we apply \puc~\cite{nouri2023puritychecker}~to select only the commits containing pure refactoring. 
Since we want to evaluate the functional correctness of the generated refactoring by running the tests, we compile every pure refactoring commit and its parent commit (to ensure there were no compilation or test failures before refactoring), selecting only the commits that could be successfully compiled and pass the tests.  


We analyze test coverage to verify whether the tests cover the refactored code (i.e., it is testing the refactored code's functional behavior). 
We execute the test cases using \textit{Jacoco} \cite{jacoco} to collect code coverage information and filter out the commits where the refactoring changes have no coverage.  
Finally, we verify the existence of target classes for the \textit{Move Method}, \textit{Extract and Move Method}, and \textit{Move and Inline Method} refactorings. This step was necessary because the \textit{Move Method} operation may move a method to newly created classes, and it is difficult for \tool to predict the newly created classes. 
After applying all the above data selection steps, we identified 703 pure refactorings across the 10 Java projects.

\noindent{\textbf{Evaluation Metrics}.} We evaluate the refactored code along two dimensions: functional correctness and human-likeness. For functional correctness, we assess the code using 1) compilation success, 2) test pass, and 3) \rfm verification (i.e., \rfm detects that the refactoring activity indeed happened). Specifically, we integrate the generated refactored code into the project. We then compile it and execute the tests to verify if the builds are successful. 
Because of LLMs' hallucination issues \cite{DBLP:journals/corr/abs-2311-05232}, the generated code may pass the test cases but do not accurately perform the intended refactoring. Hence, we verify whether the generated code contains the target refactoring using \rfm~\cite{tsantalis2018accurate}. 


Even though we give a target method as input to \tool, it still needs to find the specific part of the code that can be refactored. Hence, we evaluate the human-likeness of \tool's generated code to compare its refactoring decisions with that of developers. We employ the CodeBLEU metric \cite{ren2020codebleu} and Abstract Syntax Tree (AST) Diff Precision and Recall \cite{Alikhanifard:TOSEM:2024:RefactoringMiner3.0} to measure the difference. CodeBLEU evaluates the grammatical and semantic consistency between human-written refactored code and \tool-generated code. 
AST Diff represents a set of mappings that capture code changes between the original and refactored code. Each mapping consists of a pair of matched AST nodes from the diff between the original and refactored versions. These mappings are obtained using \rfm. To evaluate structural similarity, we compare the mappings produced by \tool's refactored code with those of the developer-written code. The number of \tool's mappings that match the developer-written mappings is treated as true positives (TP). Precision is calculated as the ratio of TP to all mappings produced by \tool, indicating how accurate \tool's mappings are. Recall is the ratio of TP to all developer-written mappings, reflecting how much of the developer’s refactoring was successfully captured.
The values for CodeBLEU and AST Precision/Recall range from 0 to 1, where 1 means a perfect match. 

\begin{table*}[]
    \centering
    \caption{Refactoring results of \raw and \tool. The table presents the number of refactorings to perform, compile-and-test success rates, refactoring verification (\textit{RM Verification}), and code similarity metrics with human-written refactorings (\textit{Code BLEU} and \textit{AST Precision/Recall}).  \textbf{Successful Refactoring} refers to the number of refactorings that compile, pass tests, and are verified by \rfm. We compute the average for Code BLEU and AST Precision/Recall, and total for all other fields. 
}
    \setlength{\tabcolsep}{2pt}
\resizebox{\textwidth}{!}{
    \begin{tabular}
    {ll|r|rrrrr||r|rrrrrr}
    \toprule
        \multirow{2}{*}{\textbf{Approach}} & \multirow{2}{*}{\textbf{Project}} & {\textbf{\# Pure}}  &  \textbf{Compile\&Test} & \textbf{RM} & \textbf{Code} & \textbf{AST} & \textbf{AST} & \textbf{Successful} & \textbf{Extract}  & \textbf{Inline}  & \textbf{Move}  & \textbf{Extract And}  & \textbf{Move And}  & \textbf{Move And}\\
        & & \textbf{Refactoring} & \textbf{Success} & \textbf{Verification} & \textbf{BLEU} & \textbf{Precision} & \textbf{Recall} &\textbf{Refactoring} &\textbf{Method} &\textbf{Method} &\textbf{Method} &\textbf{Move Method} &\textbf{Rename Method} &\textbf{Inline Method} \\
        
    \midrule
        \multirow{11}{*}{\raw} 
        & checkstyle & 91 & 14 & 9 & 0.667 & 0.603 & 0.233 & 6  & 4& 2& 0& 0& 0 & 0\\
       & pmd & 125 & 35 & 30 & 0.502 & 0.791 & 0.338 & 19  & 17& 2& 0& 0& 0 & 0\\
        & commons-lang & 59 & 2 & 13 & 0.640 & 0.67 & 0.363 & 2  & 2& 0& 0& 0& 0 & 0\\
        & hibernate-search & 89 & 14 & 27 & 0.368 & 0.792 & 0.632 & 11  & 5& 1& 0& 0& 0 & 5\\
        & junit4 & 18 & 10 & 10 & 0.486 & 0.856 & 0.859 & 9  & 8& 1& 0& 0& 0 & 0\\
        & commons-io & 93 & 8 & 22 & 0.773 & 0.804 & 0.95 & 6  & 5& 1& 0& 0& 0 & 0\\
        & javaparser & 56 & 11 & 11 & 0.441 & 0.777 & 0.96 & 4  & 2& 1& 0& 0& 0 & 1 \\
        & junit5 & 105 & 2 & 5 & 0 & 0 & 0 & 0  & 0& 0& 0& 0& 0 & 0\\
        & hibernate-orm & 63 & 17 & 2 & 0.263 & 0.756 & 0.386 & 2  & 2& 0& 0& 0& 0 & 0\\
        & mockito & 4 & 2 & 2 & 0.678 & 0.659 & 0.746 & 2  & 2& 0& 0& 0& 0 & 0\\
        \cmidrule(lr){2-15}
        & \textbf{Total/Average} & 703 & 100 & 146 &  0.517 & 0.773 & 0.574 & 61 (8.7\%)  & 47& 8& 0& 0& 0 & 6\\
    \midrule
        \multirow{11}{*}{\tool}
         &	checkstyle	&	91	&	90	&	86	&	0.624	&	0.514	&	0.501	&	85	&	31	&	4	&	9	&	39	&	1	&	0	\\
&	pmd	&	125	&	119	&	108	&	0.676	&	0.725	&	0.766	&	106	&	49	&	2	&	28	&	24	&	3	&	0	\\
&	commons-lang	&	59	&	56	&	46	&	0.567	&	0.815	&	0.257	&	46	&	42	&	1	&	0	&	3	&	0	&	0	\\
&	hibernate-search	&	89	&	81	&	82	&	0.538	&	0.929	&	0.734	&	74	&	35	&	10	&	6	&	13	&	9	&	1	\\
&	junit4	&	18	&	13	&	15	&	0.61	&	0.879	&	0.676	&	12	&	9	&	1	&	2	&	0	&	0	&	0	\\
&	commons-io	&	93	&	87	&	81	&	0.623	&	0.873	&	0.662	&	80	&	62	&	2	&	9	&	7	&	0	&	0	\\
&	javaparser	&	56	&	51	&	51	&	0.645	&	0.859	&	0.681	&	46	&	31	&	1	&	7	&	6	&	0	&	1	\\
&	junit5	&	105	&	79	&	76	&	0.852	&	0.814	&	0.787	&	74	&	9	&	0	&	48	&	15	&	2	&	0	\\
&	hbernate-orm	&	63	&	56	&	55	&	0.524	&	0.805	&	0.474	&	55	&	46	&	1	&	0	&	7	&	0	&	1	\\
&	mockito	&	4	&	4	&	4	&	0.754	&	0.861	&	0.736	&	4	&	3	&	0	&	0	&	1	&	0	&	0	\\
        \cmidrule(lr){2-15}
         &\textbf{Total/Average} &	703	&	636	&	604	&	0.64	&	0.781	&	0.635	&	582 (82.8\%)	&	317	&	22	&	109	&	115	&	15	&	3	\\
    \bottomrule \\
    \end{tabular}
}
    \label{tab:rq1_table}
\end{table*}

\noindent{\textbf{Environment}.} We selected OpenAI's ChatGPT model \cite{achiam2023gpt} for our experiment due to its popularity and ease of integration through the OpenAI-API. Specifically, we utilized the \textit{gpt-4o-mini-2024-07-18} version, as it offers a balance of affordability and strong performance. 
We implemented \tool using version 0.2.22 of \textit{LangGraph}~\cite{langgraph} and various static analysis tools that we implemented using a combination of APIs from \rfm, a modified version of \rfm, and Eclipse JDT~\cite{eclipseJDT}. 
On average, one complete refactored code generation (from querying the database, code generation, and test execution to fixing compilation and test failures) takes less than a minute on a Linux machine (Intel® Core™ i9-9900K CPU @ 3.60GHz, 64GB Memory), costing less than \$0.10. 

\subsection*{\textbf{RQ1: How effective is \tool in refactoring code?}}
\noindent{\textbf{\uline{Motivation}}.} In this RQ, we evaluate \tool's generated refactored code along two dimensions: functional correctness and human-likeness. We also analyze \tool's performance across different refactoring types. This RQ offers insights into how effective \tool is at performing refactoring tasks and the specific types of refactoring tasks where \tool is most effective.

\noindent{\textbf{\uline{Approach}}.} We evaluate \tool using the 703 pure refactoring commits collected from the 10 studied Java projects. 
First, we use {\tt git checkout} on the commit \textit{before} each pure refactoring commit, allowing us to extract the original code before refactoring operations. The code repository is then fed into \tool to generate the refactored code. 
For comparison, we include \raw as the baselines (it uses the same LLM as \tool). \raw directly sends a simple prompt to the LLM to perform code refactoring. 
We input \raw with basic code information, i.e., the same information generated by the static-analysis component of \tool's Developer Agent (e.g., class content and project structure). \raw does not have the multi-agent component and is not prompted with few-shot examples retrieved from RAG. The complete prompt can be found online~\cite{muarf_data_code}.

\noindent{\textbf{\uline{Result}}.} 
\textbf{\textit{\tool successfully generated 582/703 (82.8\%) of the refactored code that is compilable, passed all the tests, and verified by \rfm, while \raw could only generate 61/703 (8.7\%) successfully.}}  
As shown in Table \ref{tab:rq1_table}, \tool can generate significantly more refactored code than \raw. 
Of the 703 refactorings, 636 generated by \tool successfully compiled and passed the test cases, and 604 refactorings were further verified by \rfm as true refactoring operations. In contrast, only 100 refactorings generated by \raw can compile and pass the test cases, yet only 61 were verified by \rfm. Note that there can be some generated refactored code that is verified by \rfm but does not pass compilation/tests, or vice versa, so the total successful refactoring is 582.

\raw has difficulties in generating \textit{Move Method}, \textit{Extract and Move Method}, and \textit{Move and Rename Method}, where it cannot generate any refactoring. When doing these refactorings, \raw always ignores the project structure information in the prompt and cannot move the method to the correct class. In contrast, \tool has the Reviewer Agent that gives feedback on the refactoring verification to guide the Developer Agent to perform the Move operation. 
Nevertheless, even for refactorings that do not require repository or class structures (i.e., \textit{Extract Method} and \textit{Inline Method}), \tool achieves a much higher success rate (317 v.s. 47 and 22 v.s. 8, respectively).


\begin{table}
    \caption{The number of generated refactored code (compilable and pass all test cases) that is identical to that of developer's.}
    \label{tab:rq1_same}
    \centering
    \scalebox{0.72}{
    \setlength{\tabcolsep}{0.9cm}

        \begin{tabular}{lrr}		
            \hline
            \textbf{Refactoring Type} & \textbf{\raw} & \textbf{\tool} \\
            \hline
Extract Method 	&	4 &11	\\
Inline Method	&	4	&	8		\\
Move Method&	0	&	84		\\
Extract And Move Method	&	0	&		2	\\
Move And Rename Method&	0	&	0		\\
Move And Inline Method	&	0	&	0	\\
    \hline
\textbf{Total}	&	8	&	105	\\	
\hline
        \end{tabular}
        }
       \vspace{-3mm}
\end{table}

\noindent\textbf{\textit{\tool outperforms \raw in code similarity, producing refactored code more similar to humans.}} Across all successful refactorings, \tool achieved a CodeBLEU score of 0.640, compared to \raw’s 0.517, showcasing \tool's ability to generate code that closely aligns with human-written refactorings. Regarding structural accuracy, \tool achieved an AST Diff precision of 0.781, surpassing \raw’s 0.773, while its AST Diff recall reached 0.635, notably higher than \raw’s 0.574. 
These results indicate that \tool's generated code is more similar and aligns better with the structural transformation of developer-written refactoring. 

\noindent{\textbf{\textit{\tool's results are closer to developers' decisions, where 18\% (105/582) of \tool's generated refactored code is identical to developer's refactoring, compared to \raw's 13.1\% (8/61). }}} 
We further analyze the distribution of refactored code identical to the developer's implementation across different refactoring types. \tool correctly generated 84 \textit{Move Method} refactorings, whereas \raw failed to produce any valid refactorings for this category. Additionally, \tool applied 11 \textit{Extract Method} and eight \textit{Inline Method} that were the same as developers' refactoring, while \raw only managed four for both types. \tool was also able to generate two composite refactorings (\textit{Extract and Move Method}) that were identical to those of developers. 
In short, \tool's results match closer to developers' refactoring decisions, likely due to its retrieval-augmented generation (RAG) component, which provides similar past refactorings as few-shot examples.  

\rqboxc{\tool outperforms \raw in both functional correctness and human-likeness, successfully generating 582/703 (82.8\%) compilable and test-passing refactorings, compared to 61/703 (8.7\%) for \raw. \tool also has a higher CodeBLEU score, AST Diff precision and recall, with 18\% (105/582) of its refactorings identical to developers', versus 13.1\% (8/61) for \raw.}


\begin{table}
    \centering
    \caption{The number of successful \textit{Extract Method} refactorings by \emas and \tool-3.5-turbo. }
    \setlength{\tabcolsep}{4mm}
\scalebox{0.72}{
    \begin{tabular}{l|r|r|r}
    \toprule
         {\textbf{Project}} & {\textbf{\# Extract Method}}  & \textbf{\emas} & \textbf{\tool-3.5-turbo}\\

    \midrule
         checkstyle & 34 & 14 & 34\\
         pmd & 55 &  24 &48\\
         commons-lang & 54  & 18 &41\\
         hibernate-search & 28  & 25 &27\\
         junit4 & 11 & 3 &9\\
         commons-io & 68  & 33 & 49\\
         javaparser & 35 & 22 &29\\
         junit5 & 9 & 6 & 6\\
         hibernate-orm & 52 & 39 &40\\
         mockito & 3 & 1 & 1\\
        \midrule
        \textbf{Total} & 359 & 185 &277\\
    
    \bottomrule 
    \end{tabular}
}
    \vspace{1mm}
    \label{tab:rq2_table}
\end{table}

\subsection*{\textbf{RQ2: How does \tool compare to IntelliJ's LLM-based refactoring tool?}}

\noindent{\textbf{\uline{Motivation}}.}We compare the code generated by \tool to the code produced by IntelliJ's LLM-based refactoring tool. 
IntelliJ IDEA \cite{Intellij_IDEA} provides a plugin, called \emas \cite{DBLP:conf/sigsoft/PomianBDKBSBD24,pomiannext}, which uses LLM for one specific type of refactoring, i.e., \textit{Extract Method}. 
\emas utilizes in-context learning, providing all necessary instructions within the prompt, including the task definition and relevant contextual information. 
The input for \emas is the method to be refactored, and it outputs a list of suggestions that include the start and end lines to be extracted, along with the new method's name. To ensure the suggestions are valid, \emas uses IntelliJ’s static analysis abilities to filter out suggestions that would cause compilation errors. Once valid suggestions are identified, \emas applies the refactorings via the IntelliJ IDEA API based on the AST. Given its superior performance compared to tools like JDeodorant~\cite{DBLP:conf/icse/MazinanianTSV16} and GEMS~\cite{8109070}, we selected \emas as our comparison baseline.


\noindent{\textbf{\uline{Approach}}.} 
We use the latest version of \emas 0.7.5 \cite{EM-Assist0.7.5} to perform \textit{Extract Method} refactoring. This version of \emas utilizes gpt-3.5-turbo-0125 for refactoring. Since \emas 0.7.5 is fully integrated into IntelliJ IDEA and lacks an interface to change the LLM version, we also used the same version of GPT model (i.e., gpt-3.5-turbo-0125) to run \tool on all \textit{Extract Method} refactorings. 

\noindent{\textbf{\uline{Result}}.} \textbf{\textit{\tool (3.5-turbo) is able to refactor 77.1\% (277/359) of the Extract Method refactoring, while \emas can only refactor 51.5\% (185/359), providing almost 50\% improvement.}}
Table \ref{tab:rq2_table} shows the result of Extract Method Refactoring of \emas and \tool.
\emas refactors 185 out of 359 methods successfully. In comparison, \tool outperforms \emas by successfully refactoring 277 methods. Among the refactorings performed, 142 methods were successfully refactored by both tools. The finding shows that \tool is also complementary to \emas, as it successfully handled a significantly different subset of refactoring cases and demonstrates the potential to combine both techniques. 
We also see a decrease when changing \tool's underlying LLM from GPT-4o-mini to 3.5-turbo, where the number of successfully refactored Extract Method decreased from 317 (Table~\ref{tab:rq1_table}) to 277 (12.6\% decrease). This performance drop shows the impact of the underlying LLM, but the overall result is still promising. 

\rqboxc{\tool (3.5-turbo) significantly outperforms \emas in \textit{Extract Method} refactoring, achieving a 50\% improvement by successfully refactoring 277 methods compared to \emas's 185.}
\begin{figure*}[t]
  \centering
  \scalebox{1}{
  \includegraphics[width=\textwidth]{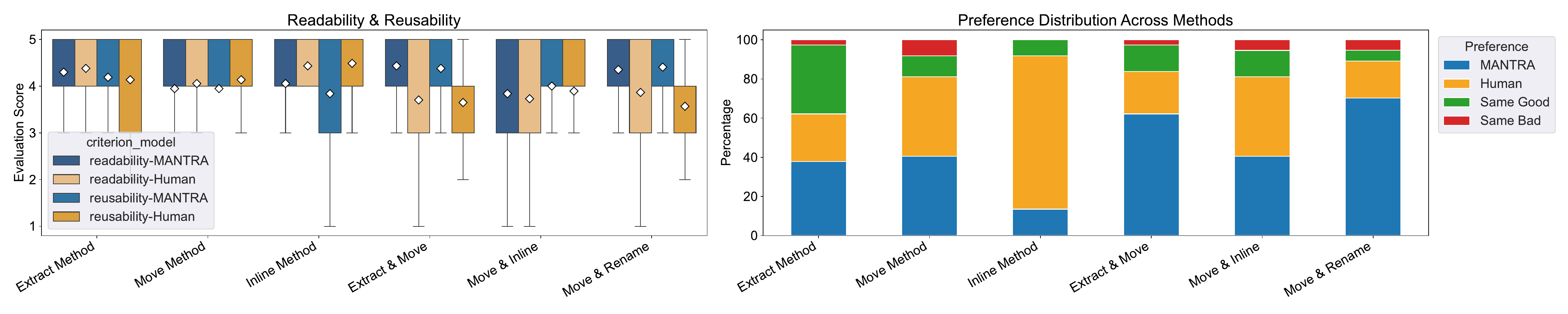}}
  \caption{Left panel: Boxplots depicting the readability and reusability scores from the questionnaire, comparing \tool-generated code with human-written code. White markers indicate the mean score for each refactoring category.
Right panel: A visualization of participants’ preferences regarding which code they favor.}
  \label{fig:rq3}
  \vspace{1mm}
\end{figure*}

\subsection*{\textbf{RQ3: How does MUARF-generated refactored code compare to human-written code?}}

\noindent{\textbf{\uline{Motivation}}.} In prior RQs, we show that \tool can successfully generate many refactorings. However, it is also important that the refactored code is  understandable and aligned with human coding practices.  
Hence, in this RQ, we conduct a user study to compare \tool's generated and human-written refactored code. 

\noindent{\textbf{\uline{Approach}}.} We randomly select 12 refactorings from the studied projects used, composing two refactorings for each of the six refactoring types. 
For each sampled refactoring, we prepare the 1) code before refactoring, 2) refactored code generated by \tool, and 3) developer's refactored code. 
To increase the sample richness and reduce the developer's evaluation time, we divide these 12 samples into two separate survey questionnaires
~\cite{refactoring_survey_1, refactoring_survey_2}, each questionnaire covers samples from the six refactoring types. 
For each sample, the participants compare two code snippets (\tool-generated and developer-written). We follow prior studies~\cite{abid202030, alomar2020developers, moser2006does, tashtoush2013impact} 
and ask the participants to assess the code's readability and reusability (on a scale from one to five, where five means highly readable or reusable) and selecting the one they find more intuitive and well-structured. For readability, we ask the participants how readable is the refactored code. For reusability, we ask the participants how easy it is to reuse or extend the refactored code.  
To avoid biases, \textit{\textbf{we do not specify which one is written by humans or generated by LLM until the participants complete the survey}}. We then ask for their opinion in free-form text after revealing this information. 


\noindent{\textbf{\uline{Result}}.} 
We shared the questionnaires through social media, and in total, we collected 37 responses for the two questionnaires (20 and 17, respectively). We combine the results from the two questionnaires and present the results below. Overall, the participants have programming experience ranging from 1 year (5.4\%) to over 5 years (54.1\%), and over 45\% of the participants use Java as their primary programming language.

\noindent\textbf{\textit{On average, the participants find that \tool-generated code has similar  readability and reusability compared to human-written code.}} As shown in Figure~\ref{fig:rq3}, LLM-generated code achieves average readability and reusability scores of 4.15 and 4.13, respectively, compared to human-written code, which scored 4.02 and 3.97. We further applied a student's t-test, and we did not find a statistically significant difference. 
This finding indicates, when considering all refactoring types, the participants find \tool's generated code has similar readability and reusability compared to human-written code. 

However, \textbf{\textit{for specific types of refactoring (i.e., \textit{Extract \& Move} and \textit{Move \& Rename}), \tool-generated code shows roughly a 20\% improvement in readability and reusability, and it is preferred 185\% more than human-written code}}, with statistically significant difference (p-value $<$ 0.001).  
In contrast, human developers achieve higher scores for \textit{Inline Method}, an average of 8.5\% higher in readability and 14.5\% in reusability, and is 439\% more preferred than \tool-generated refactored code (statistically significant with p-value $<$ 0.05). Although human developers achieve slightly higher average readability and reusability scores for \textit{Move and Inline Method}, the difference is not statistically significant. 

\begin{figure}
  \centering
  \includegraphics[width=0.5\textwidth]{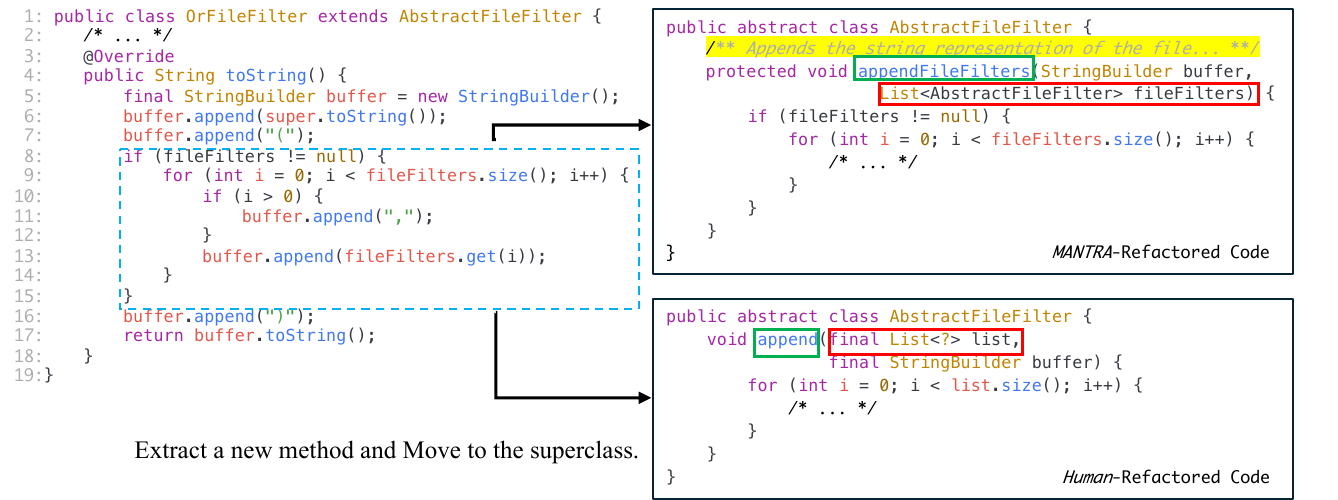}
  \caption{An example illustrating how \tool and human developers implemented the \textit{Extract \& Move} refactoring. }
  \label{fig:rq3_example}
  \vspace{1mm}
\end{figure}

Figure~\ref{fig:rq3_example} shows an example where the participants prefer the \tool-generated code (readability and reusability score of 4.50 and 4.35, compared to 3.65 and 3.75 for human-refactored code, respectively). Both \tool and a human developer performed \textit{Extract and Move} refactoring by extracting the same code snippet into the same superclass. However, they differ in the method names, comments, and parameters type. 
One participant says that ``the \textit{[LLM-generated code] is clearly easier to understand. From the comments and names, I can guess the functionality of the code... [Human-written code] is likely a very generic skeleton code.}''. 

Overall, we find a trend based on the participants' scores and responses: \textbf{\textit{\tool-generated code typically includes detailed comments and refactoring that more closely aligns with its intended purpose.}} For instance, \tool tends to use more descriptive method names that improve clarity. For instance, one participant mentioned that ``\textit{[LLM-generated code] is more structured and clear, and it seems more detailed and easier to understand.}''. 
In comparison, developer-refactored code sometimes improves code readability when doing specific refactoring types, especially during \textit{Inline Method}. For example, in one of the \textit{Inline Method} code snippets in the questionnaire, the developer improved the code during refactoring, while \tool's generated refactored code simply moved the code directly. 
A participant mentioned: ``\textit{[LLM-generated code] added two lines of code instead of one, making it harder to maintain in the future.}'' 
The study shows that participants generally prefer \tool-generated code for its clarity and detailed comments. However, in some cases, such as \textit{Inline Method}, human developers tend to write more readable and maintainable code by making additional improvements beyond direct refactoring.




\rqboxc{The participants find that \tool-generated code has similar readability and reusability compared to human-written code. \tool tends to perform better in \textit{Extract \& Move} and \textit{Move \& Rename} due to its clear comments and better naming. For \textit{Inline Method}, human developers often write more readable code by making additional improvements beyond direct refactoring. }

\subsection*{\textbf{RQ4: What is the contribution of each component in \tool?}}

\noindent{\textbf{\uline{Motivation}}.} In this RQ, we conduct an ablation study to examine the contribution of each component to the overall effectiveness of \tool. The results will highlight the importance of each component, inspiring future research on adapting them for related tasks. 


\noindent{\textbf{\uline{Approach}}.} Our ablation study examines three key components: RAG, the Reviewer Agent, and the Repair Agent. We define three configurable models to evaluate the impact of key components in \tool: \tool\textsubscript{\textit{w/o RAG}}, \tool\textsubscript{\textit{w/o Reviewer}}, and \tool\textsubscript{\textit{w/o Repair}}. 
Each of the configure removes the corresponding key component, which allows us to 
assess the contribution of each component to \tool’s overall performance.

\begin{figure}
  \centering
\scalebox{0.9}{
  \includegraphics[width=1\columnwidth]{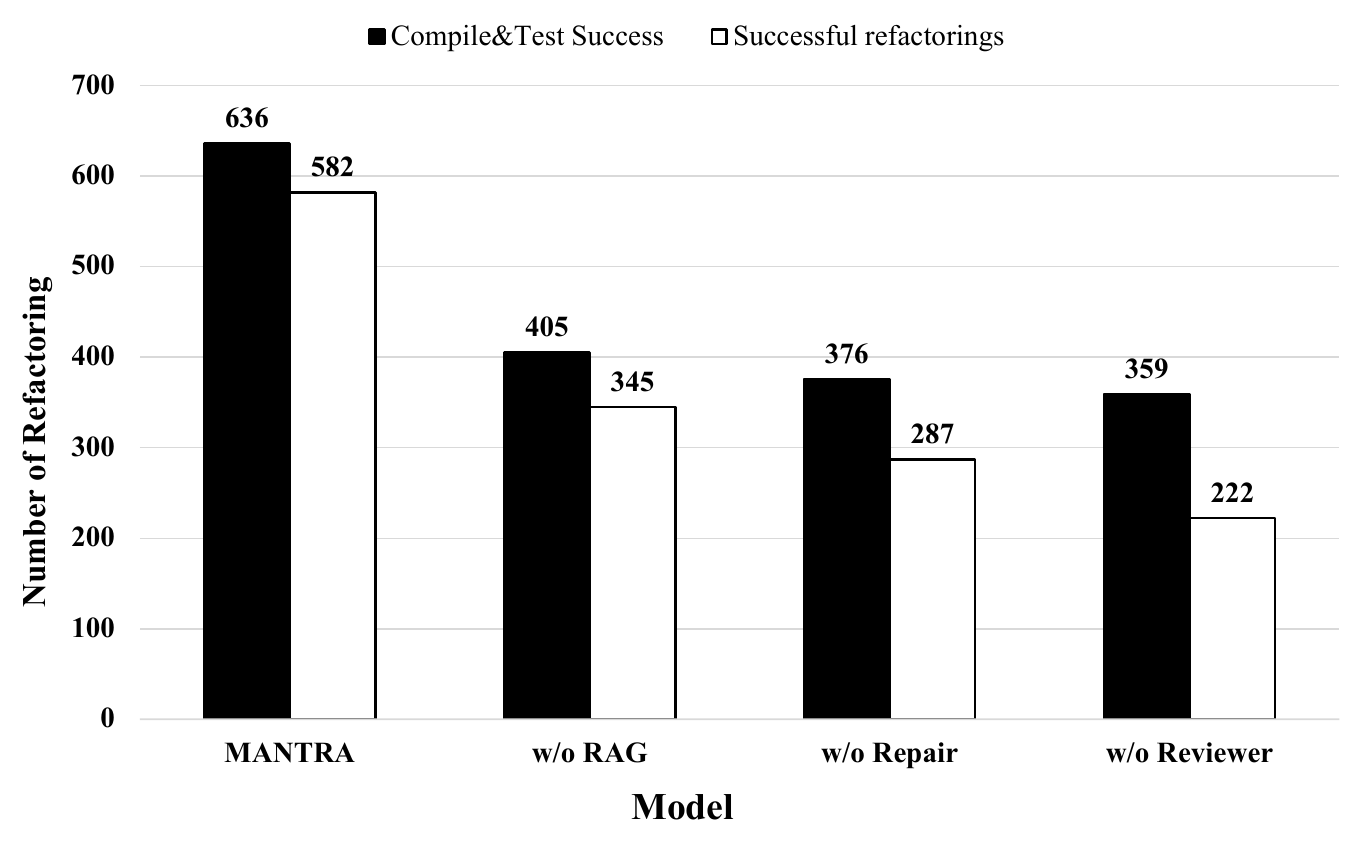}}
  \caption{Contribution of each component in \tool. \textit{Compile\&Test Success} shows the number of generated code that compiles and passes all tests. \textit{Successful refacotorings} means the number of verified code that contains the specific refactoring.  }
  \label{fig:rq2_contribution_metrics}
\end{figure}

\noindent{\textbf{\uline{Result}}.} \textbf{\textit{Removing a component reduces the number of successful refactoring from 582 to 222--345 (40.7\%--61.9\% decrease), highlighting their contribution to \tool's ability.}} 
Figure~\ref{fig:rq2_contribution_metrics} shows the contribution of each component. 
Removing the RAG component alone reduces the number of successfully compiled/tested refactored code and the number of successful refactorings to 405 and 345 (36.3\% and 40.7\% decrease), respectively. The findings show that a high-quality database for RAG has a non-negligible impact on the generated refactored code. Similarly, removing the Repair Agent significantly reduces the number of successfully compiled/tested refactored code from 636 to 376 (40.9\% decrease), as the Repair Agent is responsible for fixing compilation issues and test failures. Removing the Repair Agent also reduces the number of successful refactoring from 582 to 287 (50.7\% decrease), as the final-stage repair process plays a crucial role in finalizing the refactoring changes. Among the three components, removing the Reviewer Agent has the most impact on the number of generated refactored that pass compilation/test (decrease from 636 to 359) and the number of successful refactoring (decrease from 582 to 222). The Reviewer Agent leverages traditional tools to provide feedback in the refactoring process. Our result highlights that without feedback from external tools such as \rfm, \tool encounters challenges in generating valid refactored code. Our finding also shows a promising direction in combining traditional software engineering tools to guide LLMs in producing better results.

\rqboxc{Each component of \tool plays a crucial role in ensuring successful refactorings, with the Reviewer Agent having the most significant impact by leveraging external tools to validate and refine refactored code. The findings also highlight the importance of integrating traditional software engineering tools with LLM-based approaches, as they provide essential feedback that improves LLM's results.}


\section{THREATS TO VALIDITY} \label{sec:threats}

\noindent{\textbf{Internal validity}.} Due to the generative nature of LLMs, their responses may vary across different runs and model versions. In our experiments, we set the temperature value to 0 to reduce variability in the result. 
We used LLMs from OpenAI (i.e., 4o-mini and 3.5-turbo) for our experiment. Future studies are needed to study the impact of LLMs on generating the refactored code. 


\noindent{\textbf{External validity}.} We focused on method-level refactorings due to their popularity \cite{Negara:2013, kim2014empirical}. Although we included both straightforward and compound refactorings, the results may not generalize to other types of refactoring, such as class-level. Such refactorings are less common and often involve other code changes (e.g., bug fixes) \cite{dipenta2020relationshiprefactoringactionsbugs}, making data collection difficult. Further research is needed to assess \tool's effectiveness in broader refactoring scenarios. We focused on Java since it has extensive literature on refactoring-related research. Future work should evaluate \tool across multiple languages.


\noindent{\textbf{Construct validity}.} We use CodeBLEU and AST Precision/Recall to evaluate the similarity between \tool-generated and developer-written refactoring. However, although informative, these metrics may still miss some differences in the code. Therefore, we conducted a user study to compare \tool-generated and developer-written code. While we gathered feedback from 37 developers with different experience levels, some findings can be subjective. To avoid biases, we do not tell the participants which code is refactored by \tool or human developers until they finish the questionnaire.  

\section{Conclusion} \label{sec:conclusion}
In this paper, we introduced \tool, an end-to-end LLM agent-based solution for automated method-level code refactoring. 
By leveraging Context-Aware Retrieval-Augmented Generation, Multi-Agent Collaboration, and Verbal Reinforcement Learning, \tool generates human-like refactored code while ensuring correctness and readability. Our evaluation on 703 real-world refactorings across 10 diverse Java projects demonstrates that \tool significantly outperforms LLM-based refactoring baseline by achieving an 82.8\% success rate in generating compilable and test-passing code—far surpassing. It also has a 50\% improvement over IntelliJ’s LLM-based tool (\emas). Furthermore, our user study with 37 developers reveals that \tool-refactored code is as readable and reusable as human-written code, with better code for some specific refactoring types. 
In short, our findings highlight the potential of LLM-based refactoring tools in automating software maintenance. 

\balance
\bibliographystyle{plainnat}
\bibliography{custom}

\end{document}